\documentclass[letter]{JHEP3}
\usepackage{graphicx}
\def\pr{{^\prime}}
\def\dpr{{^{\prime\prime}}}
\def\qp{ {\bf q}_\perp }
\def\pp{ {\bf p}_\perp }
\def\lp{ {\bf l}_\perp }
\def\kp{ {\bf k}_\perp }
\newcommand{\kpn}[1]{ {\bf k}_{#1\perp} }
\title{Three-particle correlation from glasma flux tubes}
\author{Kevin Dusling\\Physics Department\\Building 510A\\Brookhaven National Laboratory\\Upton, NY-11973, USA\\Email: \email{kdusling@quark.phy.bnl.gov}}
\author{Daniel Fern\'andez-Fraile\\ Departamento de F\'isica Te\'orica II\\Univ. Complutense, 28040 Madrid, Spain\\Email: \email{danfer@fis.ucm.es}}
\author{Raju Venugopalan\\Physics Department\\Building 510A\\Brookhaven National Laboratory\\Upton, NY-11973, USA\\Email: \email{raju@bnl.gov}}

\abstract{We compute three particle correlations in the Glasma flux tube model of high energy heavy ion collisions. We obtain a simple geometrical 
picture of these correlations; when convoluted with final state radial flow, it results in distinct predictions for the near side three particle correlation in 
central heavy ion collisions.}

\begin{document}

\section{Introduction}

Recent experiments at RHIC by the STAR~\cite{STAR1}, PHENIX~\cite{PHENIX} and PHOBOS~\cite{Alver:2008gk} collaborations have 
demonstrated the existence of striking ``ridge" like structures in the near side two particle correlation spectrum. In near side events with 
prominent jet like structures, the spectrum of associated particles is collimated in the azimuthal separation $\Delta \Phi$ relative to the jet and 
shows a nearly constant amplitude in the strength of the pseudo-rapidity correlation $\Delta \eta$ up to $\Delta \eta\sim 1.5$~\cite{STAR2}.
This collimated correlation persists up to $\Delta \eta ~\sim 4$~\cite{Molnar:2007wy,Wenger:2008ts}. An important feature of ridge correlations is that the above 
described structure is also seen in two particle correlations without a jet trigger~\cite{STAR3}. In such events, a sharp rise in the amplitude of ridge events 
is seen~\cite{STAR3} in going from peripheral to central collisions.  A number of theoretical models have been put forth to 
explain these ridge correlations~\cite{Models}. 

In Ref.~\cite{Dumitru:2008wn} it was suggested that the ridge can be explained by the radial flow of approximately boost invariant Glasma flux tubes; gluons are produced with isotropic azimuthal distributions in the rest frame of each of these flux tubes. It was shown subsequently that STAR results on the centrality dependence of the amplitude of the ridge and of its $\Delta \Phi$ width, for two different energies, could be quantitatively reproduced in this framework with only one free parameter~\cite{Gavin:2008ev}. In this paper, we will compute three particle correlations in the Glasma flux tube picture and predict the corresponding near side ridge structure of these correlations. Our work is motivated by first  studies of three particle near side correlations at RHIC~\cite{Netrakanti:2008jw}.

The Glasma flux tube picture arises from first principles in the Color Glass Condensate (CGC)~\cite{CGC,MV} approach to multi-particle production in high energy heavy ion collisions. To all orders in perturbation theory, in the leading logarithmic approximation in $x$, the $n$-gluon correlation spectrum in high energy nucleus--nucleus collisions can be expressed as~\cite{Gelis:2008rw,Gelis:2008ad} 
\begin{eqnarray}
\left< {dN_n\over d^2 p_{\perp,1} dy_1\cdots d^2 p_{\perp,n} dy_n}\right> &=& \int [d\rho_A d\rho_B] W_{y_{\rm beam}-Y} [\rho_A] W_{y_{\rm beam} +Y}[\rho_B] \nonumber\\
&\times&{dN_{\rm LO}\over d^2 p_{\perp,1}dy_1}(\rho_A,\rho_B) \cdots {dN_{\rm LO}\over d^2 p_{\perp,n} dy_n}(\rho_A,\rho_B) \, .
\label{eq:nglue}
\end{eqnarray}
Here $\rho_A$ and $\rho_B$ are the color charge densities of the two nuclei, whose distributions are determined by the universal weight functionals 
$W$, as described in the CGC framework. The weight functionals satisfy the JIMWLK renormalization group equations~\cite{JIMWLK} which determine their evolution with the rapidity $Y$ relative to the beam rapidity $Y_{\rm beam}$. This expression is valid as long as the rapidity gap between two particles is such that $|y_i -y_j| \leq 1/\alpha_S$, where $y_i$ and $y_j$ are the rapidity of the two observed particles. The rapidity $Y$ collectively denotes rapidities in the interval $y_1,\cdots y_n$. When the rapidity separation between any two gluons exceeds this value, additional gluons can be emitted between {\it tagged} gluons\footnote{This additional radiation modifies eq.~(\ref{eq:nglue}) significantly~\cite{Gelis:2008sz}. For simplicity, we will not quote the full expression here but refer the reader to Ref.~\cite{Gelis:2008sz}.}. Because our focus here will be on the STAR data, whose pseudo-rapidity coverage is limited to $\Delta \eta\sim 1.5$ units, eq.~(\ref{eq:nglue}) is the appropriate expression for our purposes. 

The leading order single particle inclusive distribution, for a fixed distribution of sources, is given by 
\begin{eqnarray}
\frac{dN_{\rm LO}}{d^2 p_\perp dy_p }(\rho_A,\rho_B) &=&\frac{1}{16\pi^3}
\lim_{x_0,y_0\to+\infty}\int d^3x\, d^3y
\;e^{ip\cdot(x-y)}
\;(\partial_x^0-iE_p)(\partial_y^0+iE_p)
\nonumber\\
&&\qquad\qquad\times\sum_{\lambda}
\epsilon_\lambda^\mu(p)\epsilon_\lambda^\nu(p)\;
A_\mu(x) (\rho_A,\rho_B) A_\nu(y)(\rho_A,\rho_B)\; .
\label{eq:single-inclusive}
\end{eqnarray}
Here, the gauge fields $A_\mu(x)$ are solutions of the classical Yang-Mills equations in the forward light cone after the nuclear collision. Analytical solutions for these gauge fields are known to leading order\footnote{For interesting recent work in analytical treatments of this problem, see Ref.~\cite{Blaizot:2008yb}.} in the sources~\cite{Kovner:1995ts,Kovchegov:1997ke}; the full solution, to all orders in $\rho_A,\rho_B$, requires a numerical computation~\cite{KNV}. In the nuclei, before the collision, the typical range of color correlations is the saturation scale $Q_S$, where 
${Q_S}^{-1} < {\Lambda_{\rm QCD}}^{-1}$. The saturation scale at a given transverse position in the nucleus depends on the two dimensional transverse projection of the nuclear matter distribution. In eq.~(\ref{eq:nglue}), it appears in the initial conditions for the $W$ functionals; the energy evolution of the 
saturation scale is determined by the renormalization group equations.   Because the saturation scales in the two nuclei are the only scales in the problem, besides the nuclear radii, the energy and centrality dependence of the inclusive observables are a consequence of the properties of the saturation scales in the nuclei. 

The expression in eq.~(\ref{eq:nglue}) is remarkable because it suggests that in a single event--corresponding to a particular configuration of color sources--the leading contribution is from $n$ tagged gluons that are produced independently. The coherence in n-particle gluon emission is instead generated by 
averaging over color sources that vary from event to event. Because the range of color correlations in the transverse plane is of order $1/Q_S$, 
this formalism suggests an intuitive picture of coherent multiparticle production as arising from fluctuations in particle production from color flux tubes of 
size $1/Q_S$ from event to event. 

The highly coherent matter produced immediately after the nuclear collision is the Glasma~\cite{LappiMcLerran,GelisVLectures}. Besides providing the underlying geometrical structure for long range rapidity correlations, the Glasma flux tubes also carry 
topological charge~\cite{KharzeevKV}, which may result in observable metastable CP-violating domains~\cite{KharzeevMcLW}. While eqs.~(\ref{eq:nglue}) and ~(\ref{eq:single-inclusive}) describe particle production from the Glasma flux tubes, they do not describe the subsequent final state interactions of the 
produced gluons. If, as widely believed, the produced matter thermalizes by final state interactions, these will not significantly alter long range rapidity correlations. However, the radial flow of this matter will have a significant effect on the observed angular correlations. This is because even particles produced isotropically in a given flux tube will be correlated by the radial outward hydrodynamic flow of the flux tubes. Elements of these ideas were 
present in the literature~\cite{Voloshin,Shuryak}-they were proposed in the Glasma flux tube picture in Ref.~\cite{Dumitru:2008wn} and developed further along with analysis of the two particle correlations measured by STAR in Ref.~\cite{Gavin:2008ev}. 

We will extend the approach outlined above to compute the three particle spectrum in nucleus-nucleus collisions in the Glasma flux tube model. The paper 
is organized as follows. In the next section, we briefly review the two particle computation and then proceed to discuss the Glasma 3-particle correlation. 
In section 3, we compute how radial flow affects the three particle correlation. In the following section, we discuss our results and predictions for experimental three particle correlations. We conclude with a brief summary. Technical details of the computation are contained in three appendices.

\section{Computing the Glasma 3-particle correlation}

Before coming to the three particle correlation let us first summarize the result of the two particle correction computed in \cite{Dumitru:2008wn}.  Many of the arguments made here will carry over to the three particle case.  We will focus in this section on the intrinsic multi-parton correlations arising from the physics of high parton densities. The effects of final state interactions, in particular the effects of radial flow, will be addressed in the next section.

\begin{figure}
\begin{center}
  \resizebox*{!}{3cm}{\includegraphics{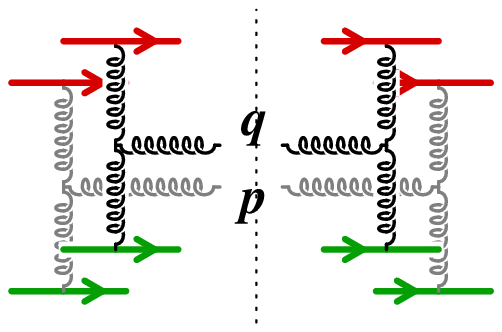}}
\end{center}
\begin{center}
  \resizebox*{!}{3cm}{\includegraphics{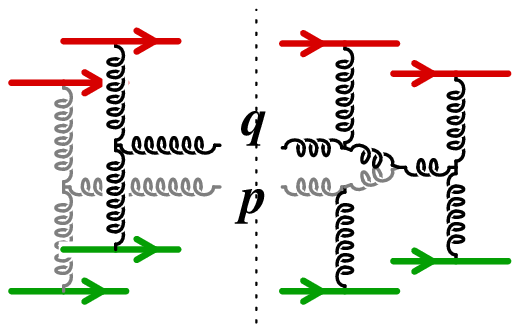}}
\end{center}
\caption{Top Figure: A classical diagram which yields a non-vanishing
  two particle correlation after averaging over the color sources.
  Bottom Figure: A contribution to the correlation function associated
  with a quantum correction to the classical field. In the strong field limit, each source goes like $1/g$ while each vertex yields a power of $g$.  One therefore finds that the bottom diagram is power suppressed relative to the top diagram by $g^2$.  Analogous arguments hold for the three particle case. }
\label{fig:illust1}
\end{figure}

For two particles having transverse momentum ${\bf p}_\perp, {\bf q}_\perp$ the correlation takes the form
\begin{eqnarray}
C_2({\bf p},{\bf q})&\equiv&\left<\frac{d^2 N_2}{dy_pd^2{\bf p}_\perp dy_qd^2{\bf q}_\perp }\right> -\left<\frac{dN}{dy_pd^2{\bf p}_\perp}\right>\left<\frac{dN}{dy_qd^2{\bf q}_\perp}\right>\,.
\label{eq:C2part}
\end{eqnarray}
Let us first start by considering a fixed configuration of color sources.  Then there will be connected and disconnected pieces
as shown for example in fig.~\ref{fig:illust1}. One might naively think the disconnected diagram would simply cancel with the uncorrelated terms in eq.~(\ref{eq:C2part}).  This is not the case because the averaging over the color sources brings about non-trivial connections between the seemingly disconnected diagrams.

Any diagram that appears to be connected for a fixed distribution of color sources is either included in the renormalization group evolution of the sources 
(if it contains as many powers of the rapidity as of $\alpha_s$) or is else suppressed~\cite{Gelis:2008rw,Gelis:2008ad}. 
The net result is that only the classical disconnected graphs contribute; all the effects of quantum evolution are hidden in the source distributions. 
We shall compute these classical contributions and evaluate the non-trivial correlations that result when they are averaged over the source distributions.
In performing the averaging, there will be trivial color correlations which will cancel with the subtractions in eqs.~(\ref{eq:C2part}).  The final result for the two particle correlation is 
\begin{eqnarray}
C_2({\bf p},{\bf q})=\kappa_2\frac{1}{S_\perp Q_S^2}\left<\frac{dN}{dy_pd^2{\bf p}_\perp}\right>\left<\frac{dN}{dy_qd^2{\bf q}_\perp}\right>\,,
\label{eq:C2part2}
\end{eqnarray}
where 
\begin{eqnarray}
\kappa_2=\frac{\pi}{(N_c^2-1)}\frac{1}{\ln\left(\frac{\pp}{Q_S}\right)\ln\left(\frac{\qp}{Q_S}\right)}\sim 0.4
\end{eqnarray}
was estimated analytically in \cite{Dumitru:2008wn}\footnote{The work \cite{Dumitru:2008wn} as well as a prior version of our work had some erroneous factors of 2 and $2\pi$ \cite{Gelis:2009wh}.  We have corrected for this in quoting the above value of $\kappa_2$.}.  It was found that the particles in a single flux tube, in the kinematic range $Q_S \ll k_\perp$, are produced isotropically in the azimuthal direction around the beam axis and two particle correlations are independent of the relative azimuthal angle between the tagged gluons. It was then conjectured that this isotropic distribution also holds for $k_\perp \leq Q_S\sim1$ GeV. Recent non-perturbative numerical simulations of Yang--Mills equations appear to confirm this conjecture~\cite{LappiSrednyakRV}. These numerical simulations also provide a more accurate determination of the constant $\kappa_2$. 

The above result has a physically intuitive meaning.  If only particles within a flux tube are correlated with each other, the  ratio of the two particle correlation divided by the product of 
the two single particle inclusive distributions is then simply proportional to the ratio of the flux tube area $Q_S^{-2}$ to the total system size area $S_\perp$. If this geometrical picture is correct, it should also hold for higher particle correlations. In particular, we will show that the three particle correlation can be expressed as
\begin{eqnarray}
C_3({\bf p},{\bf q}, {\bf l})=\kappa_3\frac{1}{S_\perp^2 Q_S^4}\left<\frac{dN}{dy_pd^2{\bf p}_\perp}\right>\left<\frac{dN}{dy_qd^2{\bf q}_\perp}\right>\left<\frac{dN}{dy_ld^2{\bf l}_\perp}\right>
\label{eq:C3r3}
\end{eqnarray}
where $\kappa_3\sim 0.3$. The three particle correlation divided by the product of three single particle inclusive distributions is simply proportional to square of the ratio of the flux tube area over the transverse area of the system.

We now turn to the explicit calculation of the three particle correlation.  We first start with the definition of the variance of the three particle multiplicity distribution, for three particles having momentum ${\bf p}$, ${\bf q}$ and ${\bf l}$.
\begin{eqnarray}
\label{eq:C3}
C_3({\bf p},{\bf q}, {\bf l})&\equiv&\left<\frac{d^3 N_3}{dy_pd^2{\bf p}_\perp dy_qd^2{\bf q}_\perp dy_ld^2{\bf l}_\perp}\right>\\
&-&\left<\frac{d^2N_2}{dy_pd^2{\bf p}_\perp dy_qd^2{\bf q}_\perp}\right>\left<\frac{dN}{dy_ld^2{\bf l}_\perp}\right>-\left<\frac{d^2N_2}{dy_qd^2{\bf q}_\perp dy_ld^2{\bf l}_\perp}\right>\left<\frac{dN}{dy_pd^2{\bf p}_\perp}\right>\nonumber\\
&-&\left<\frac{d^2N_2}{dy_pd^2{\bf p}_\perp dy_ld^2{\bf l}_\perp}\right>\left<\frac{dN}{dy_qd^2{\bf q}_\perp}\right>+2\left<\frac{dN}{dy_pd^2{\bf p}_\perp}\right>\left<\frac{dN}{dy_qd^2{\bf q}_\perp}\right>\left<\frac{dN}{dy_ld^2{\bf l}_\perp}\right>\nonumber
\end{eqnarray}
In the above expression, the angular brackets $\langle \cdots \rangle$, designate an event averaged quantity.  In our formalism this amounts to averaging over the color sources of the two nuclei.  In order to compute $C_3({\bf p},{\bf q}, {\bf l})$ we need  simply compute the connected contributions to the first term in eq.~(\ref{eq:C3}).
\begin{eqnarray}
\left<\frac{d^3 N_3}{dy_pd^2{\bf p}_\perp dy_qd^2{\bf q}_\perp dy_ld^2{\bf l}_\perp}\right>=\frac{1}{8(2\pi)^9}\sum_{a,a\pr,a\dpr,\lambda,\lambda\pr,\lambda\dpr}\left< \vert \mathcal{M}_{\lambda\lambda\pr\lambda\dpr}^{aa\pr a\dpr}({\bf p},{\bf q},{\bf l})\vert ^2\right>
\label{eq:melement}
\end{eqnarray}
The classical contribution to $C_3({\bf p},{\bf q},{\bf l})$ can be computed analytically for $Q_S \ll p_\perp,q_\perp,l_\perp$.  Just as in the case of the two particle correlations, we anticipate that the result at large transverse momentum will demonstrate key features of the correlation that 
are generic and will therefore persist at lower momentum as well, even though the overall numerical coefficient in  the result will likely differ. This 
conjecture can be confirmed by numerical solutions of the classical Yang--Mills equations; as we discussed previously, ongoing numerical work for the 
two particle correlations appear to confirm the conjecture in that case. Nevertheless, it is important to note that for large 
$p_\perp$, the present formalism may break down and other mechanisms such as, for example, beam jet induced rapidity 
correlations may become important.

The classical amplitude for the production of three gluons having momentum ${\bf p}, {\bf q},{\bf l}$ is
\begin{eqnarray}
\mathcal{M}_{\lambda\lambda\pr\lambda\dpr}^{aa\pr a\dpr}({\bf p},{\bf q},{\bf l}) = \epsilon^\lambda_{\alpha}({\bf p})\epsilon^{\lambda\pr}_\beta({\bf q})\epsilon^{\lambda\dpr}_\gamma({\bf l})p^2 q^2 l^2 A^{\alpha,a}({\bf p}) A^{\beta,a\pr}({\bf q}) A^{\gamma,a\dpr}({\bf l})
\label{eq:clamp}
\end{eqnarray}
where the $\epsilon$'s are the polarization vectors of the gluons with the polarization indices $\lambda$,$\lambda\pr$ and $\lambda\dpr$ and the 
$a$,$a\pr$,$a\dpr$ are color indices for the gauge fields.
Following the discussion in Ref.~\cite{Dumitru:2008wn} (and references therein), for large transverse momentum the classical gauge fields can be expressed as
\begin{equation}
p^2A^{\mu,a}({\bf p})=-if_{abc}g^3\int\frac{d^2\kp}{(2\pi)^2}\,L^\mu({\bf p},\kp)\frac{\tilde{\rho}^b_1(\kp)\tilde{\rho}_2^c(\pp-\kp)}{\kp^2\left(\pp-\kp\right)^2}
\end{equation}
where $f_{abc}$ are the SU(3) structure constants and $\tilde{\rho}_1,\tilde{\rho}_2$ are the Fourier transforms of the color charge densities of the two 
nuclei. Here $L^\mu$ is the well-known Lipatov vertex which is discussed further in appendix A. 
Performing the sum over polarizations\footnote{We work in Feynman gauge: $\sum_\lambda \epsilon^{*\lambda}_\mu \epsilon^{\lambda}_\nu = -g_{\mu\nu}$}, color indices and taking the modulus squared of eq.~(\ref{eq:clamp}), $dN_3$ can be expressed as
\begin{eqnarray}
\left<\frac{d^3 N_3}{dy_pd^2{\bf p}_\perp dy_qd^ 2{\bf q}_\perp dy_ld^2{\bf l}_\perp}\right>&=&\frac{1}{8(2\pi)^9}\left(-ig^3\right)^6\left( f_{abc}f_{a\pr de}f_{a\dpr fg}f_{a hi}f_{a\pr jk} f_{a\dpr lm}\right) \int\prod_{i=1}^6 \frac{d^2{\bf k}_{i\perp}}{(2\pi)^2{\bf k}_{i\perp}^2}\nonumber\\
&\times&\frac{L_\alpha({\bf p},\kpn{1}) L^\alpha({\bf p},\kpn{2})}{\left(\pp-\kpn{1}\right)^2 \left( \pp-\kpn{2}\right)^2}\nonumber\\
&\times&\frac{L_\beta({\bf q},\kpn{3}) L^\beta({\bf q},\kpn{4})}{\left(\qp-\kpn{3}\right)^2 \left( \qp-\kpn{4}\right)^2}\nonumber\\
&\times&\frac{L_\gamma({\bf l},\kpn{5}) L^\gamma({\bf l},\kpn{6})}{\left(\lp-\kpn{5}\right)^2 \left( \lp-\kpn{6}\right)^2}\nonumber\\
&\times& \mathcal{F}^{bcdefghijklm}({\bf p},{\bf q},{\bf l};\{\kpn{i}\})
\label{eq:c}
\end{eqnarray}
where
\begin{eqnarray}
\mathcal{F}^{bcdefghijklm}({\bf p},{\bf q},{\bf l};\{\kpn{i}\})\equiv\Big< \tilde{\rho}^{*h}_1(\kpn{2}) \tilde{\rho}^{*j}_1(\kpn{4}) \tilde{\rho}^{*l}_1(\kpn{6}) \tilde{\rho}^{b}_1(\kpn{1}) \tilde{\rho}^{d}_1(\kpn{3}) \tilde{\rho}^{f}_1(\kpn{5}) \nonumber\\
\times \tilde{\rho}^{*i}_2(\pp-\kpn{2}) \tilde{\rho}^{*k}_2(\qp-\kpn{4}) \tilde{\rho}^{*m}_2(\lp-\kpn{6}) \tilde{\rho}^{c}_2(\pp-\kpn{1}) \tilde{\rho}^{e}_2(\qp-\kpn{3}) \tilde{\rho}^{g}_2(\lp-\kpn{5}) \Big>\,. \nonumber\\
\label{eq:F}
\end{eqnarray}

The average in the above expression for $\mathcal{F}$ corresponds to an average over the color sources
\begin{eqnarray}
\langle\mathcal{O}\rangle=\int\left[D\rho_1 D\rho_2\right] W[\rho_1]W[\rho_2]\mathcal{O}[\rho_1,\rho_2]\, .
\end{eqnarray}
In the MV model for large nuclei~\cite{MV}, the weight functional is modelled by a local Gaussian
\begin{eqnarray}
W[\rho]\equiv \exp\left( -\int d^2 {\bf x}_\perp \frac{\rho^a({\bf x}_\perp) \rho^a({\bf x}_\perp)}{2\mu_A^2}\right)\,.
\end{eqnarray}
In the above expression $\mu_A^2$ is the color charge squared per unit area and can be expressed in terms of the saturation scale $Q_S$ as 
$Q_S \sim 0.6{\mbox{ }} g^2 \mu_A$~\cite{Lappi:2007ku}.
For a Gaussian weight functional, the source correlator in momentum space is 
\begin{eqnarray}
\langle \tilde{\rho}^{*a}(\kpn{1}) \tilde{\rho}^{b}(\kpn{2}) \rangle = (2\pi)^2\mu_A^2\, \delta^{ab}\,\delta(\kpn{1}-\kpn{2})\nonumber\\
\langle \tilde{\rho}^{a}(\kpn{1}) \tilde{\rho}^{b}(\kpn{2}) \rangle = (2\pi)^2\mu_A^2 \,\delta^{ab}\,\delta(\kpn{1}+\kpn{2})
\end{eqnarray}
At small $x$, where quantum evolution effects are large, the weight functional is given by the solution of the JIMWLK Hamiltonian~\cite{JIMWLK}.  

\begin{figure}
\centering
\includegraphics[scale=1]{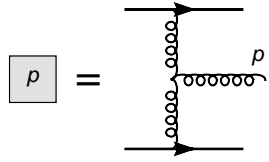}
\label{fig:defn}
\caption{Graphical notation used for an emitted gluon of momentum ${\bf p}$.}
\end{figure}

\begin{figure}
\centering
\includegraphics[scale=1]{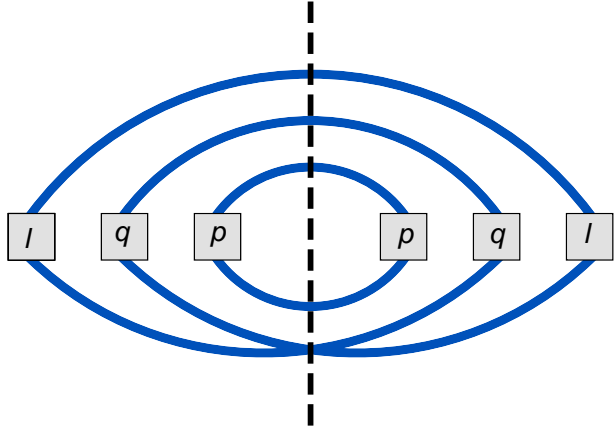}
\caption{Example of one of the twenty-seven completely disconnected diagrams.  This type of contribution to the three particle spectrum cancels in the difference of eq.~(\protect\ref{eq:C3}). }
\label{fig:disc}
\end{figure}

We can now finally discuss the evaluation of the three particle correlation.  Examining the structure of $\mathcal{F}$ in equation (\ref{eq:F}), we see that there are $15\times 15=225$ possible quadratic combinations of the $\rho_1$'s and $\rho_2$'s.  We shall now convince the reader that of these 225 possible diagrams only 16 diagrams contribute to the intrinsic three particle correlation in the limit of $Q_S \ll p_\perp,q_\perp,l_\perp$. 
In order to simplify our discussion of the diagrams, we introduce, as shown in fig.~\ref{fig:defn}, a graphical notation for an emitted gluon.  Then the contraction among the sources can be made by solid lines at the top and bottom of the boxes as shown for example in fig.~\ref{fig:disc}.

As mentioned previously, a large class of diagrams will be disconnected and will cancel with the subtractions in eq.~(\ref{eq:C3}).  An example of a completely disconnected diagram is shown in fig.~\ref{fig:disc}.  There are a total of 27 such diagrams which can be ignored.

\begin{figure}
\centering
\includegraphics[scale=1]{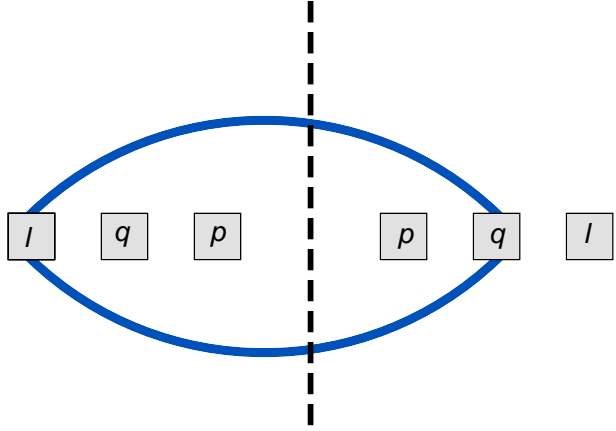}
\caption{Example of a self-connected power suppressed contribution proportional to $\delta(\lp-\qp)$.}
\label{fig:connconn}
\end{figure}

Another class of diagram which can be ignored are those which are self-connected.  An example is shown in fig.~\ref{fig:connconn}.  Regardless how one performs the remaining contractions, the result will be proportional to $\delta(\lp-\qp)$ and are power suppressed. The $\delta$ function contribution will be smeared out by final state re-scattering.  

\begin{figure}
\centering
\includegraphics[scale=1]{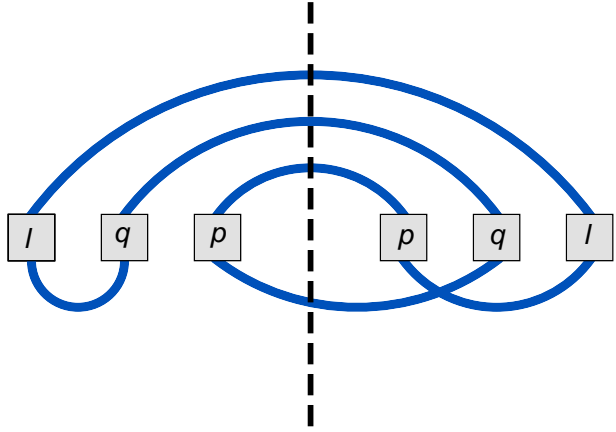}
\caption{One of the twelve diffractive diagrams that contribute to the three particle correlation.}
\label{fig:d1}
\end{figure}

\begin{figure}
\centering
\includegraphics[scale=1]{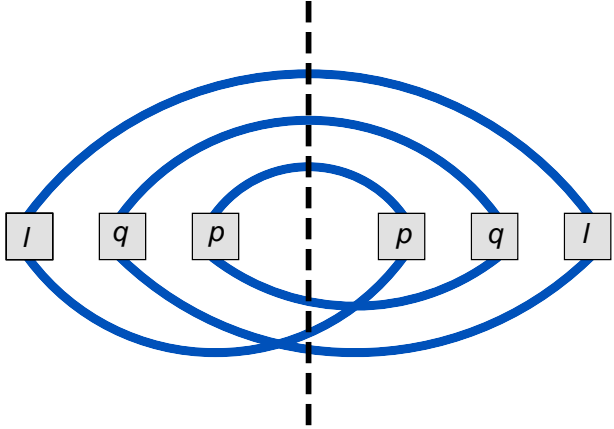}
\caption{One of the four interference diagrams that contribute to the three particle correlation.}
\label{fig:d2}
\end{figure}

There are still many diagrams remaining.  For large $\pp,\qp,\lp$ the leading contribution will come from diagrams with the minimum number of crossings in their contracting lines.  A similar observation  was made for the two particle correlation case.  In the three particle case there are sixteen diagrams of minimal crossing.  Twelve are the diffractive diagram as shown in fig.~\ref{fig:d1} and four are interference diagrams as shown in fig.~\ref{fig:d2}.  The diagrams corresponding to the topologies in fig.~\ref{fig:d1} and fig.~\ref{fig:d2} are evaluated in appendices A and B respectively.  We find that all sixteen diagrams yield the same result; the two sets of topologies can be related by simple transformations of the momentum flow in the graphs.  Multiplying either result in the appendices by 16 we find
\begin{eqnarray}
C_3({\bf p},{\bf q},{\bf l})=\frac{S_\perp}{32\pi^{11}}\frac{(g^2\mu_A)^{12}}{g^6 Q_S^4}\frac{\pi N_c^3(N_c^2-1)}{p_\perp^4 q_\perp^4 l_\perp^4}
\label{eq:C3r1}
\end{eqnarray}

It will be convenient to express the above result in terms of the inclusive single gluon spectrum.  This has been shown in \cite{Kovner:1995ts,Kovchegov:1997ke,Gunion:1981qs,Gyulassy:1997vt} to have the form
\begin{eqnarray}
\left<\frac{dN}{dy_p d^2\pp}\right>=\frac{S_\perp}{4\pi^4}\frac{(g^2\mu_A)^4}{g^2}\frac{N_c(N_c^2-1)}{p_\perp^4}\ln\left(\frac{p_\perp}{Q_S}\right)
\end{eqnarray}
Finally, up to logarithmic corrections, as anticipated at the start of this section, the  three particle correlation is
\begin{eqnarray}
C_3({\bf p},{\bf q}, {\bf l})=\kappa_3\frac{1}{S_\perp^2 Q_S^4}\left<\frac{dN}{dy_pd^2{\bf p}_\perp}\right>\left<\frac{dN}{dy_qd^2{\bf q}_\perp}\right>\left<\frac{dN}{dy_ld^2{\bf l}_\perp}\right>
\label{eq:C3r2}
\end{eqnarray}
with
\begin{eqnarray}
\kappa_3=\frac{2\pi^2}{(N_c^2-1)^2}\frac{1}{\ln\left(\frac{\pp}{Q_S}\right)\ln\left(\frac{\qp}{Q_S}\right)\ln\left(\frac{\lp}{Q_S}\right)}\sim 0.3
\end{eqnarray}
At this stage it is difficult to address the theoretical uncertainty in $\kappa_3$ because of the logarithmic contributions. 
The full result requires a numerical computation of the three particle correlation from solutions of classical Yang-Mills equations; albeit straightfoward, such computations are time consuming. As mentioned previously, because $Q_S$ and the transverse area are the only scales in the problem, we anticipate that the geometrical structure of the result will be robust, with the primary uncertainty being the value of $\kappa_3$. Finally, we note that as the number of 
participants in the collision goes as $N_{\rm part}\sim S_\perp Q_S^2$, our geometrical result predicts that $C({\bf p},{\bf q}, {\bf l})\propto 1/N_{\rm part}^2$. 

\section{Effect of radial flow}

Before we come to a general discussion on the phenomenological consequences of the above calculation, it is important to first consider the effect of transverse flow on the three particle correlation. In this section, we will compute the behavior of the three particle correlation function when subject to a boost in the transverse radial direction.  

Because the size of the flux tubes in the transverse plane is very small ($1/Q_S^2 \ll S_\perp$), the particles emitted by a Glasma flux tube experience 
a common flow velocity. In the local rest frame of the flux tube, the three particle correlation at large transverse momentum, given by the 
expression in eq.~(\ref{eq:C3r1}), is independent of both rapidity and azimuthal angle. In the absence of flow, we find a {\em flat} distribution in $\Delta \eta_{pq}$ vs. $\Delta\eta_{pl}$ as well as in the azimuthal correlation $\Delta\phi_{pq}$ vs. $\Delta\phi_{pl}$.  Even though this result is featureless, it is nonetheless highly non-trivial as 
we shall discuss in section \ref{sec:discussion}.

\begin{figure}
\centering
\includegraphics[scale=.7]{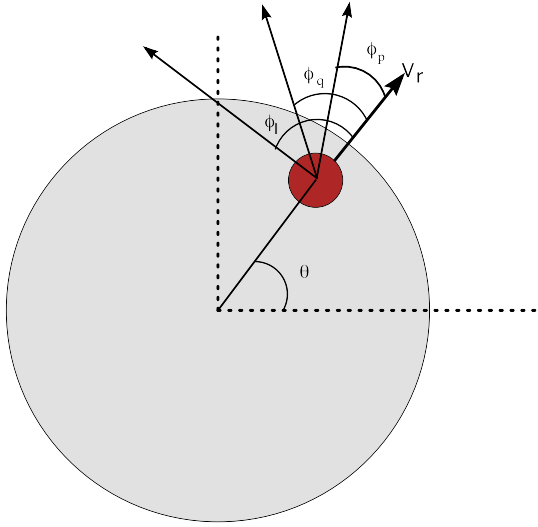}
\caption{Schematic picture of three correlated particles emitted from a flux tube of transverse size $1/Q_S^2$ and radial velocity $V_r$.  The transverse size of the collision region is defined as $S_\perp$.}
\label{fig:trans}
\end{figure}

The flat $\Delta \eta$ distribution is unaffected by the radial flow of the medium.  As we shall now describe, additional azimuthal correlations are 
generated by the flow. As done previously in \cite{Dumitru:2008wn}, we begin by introducing the particle's rapidity in the direction of transverse flow, $\zeta_{p,q,l}\equiv-\ln\left(\tan(\phi_{p,q,l}/2)\right)$, where the particle's azimuthal angle, $\phi_{p,q,l}$ is defined with respect to the radial vector pointing towards the emission point--the location of  the flux tube in the transverse plane.  Since all the particles are localized within $1/Q_S$ of each other, we expect the same radial boost for all three particles.  Fig.~\ref{fig:trans} shows the setup.

Because the three particle correlation as computed in the flux tube rest frame is independent of the azimuthal angle, the effect of the radial boost enters at the level of the Jacobian obtained when one expresses the angular distribution in terms of rapidity variables 
\begin{eqnarray}
C_3\sim \frac{d^3 N_3}{d\phi_p d\phi_q d\phi_l}=\cosh\zeta_p\cosh\zeta_q\cosh\zeta_l\frac{d^3 N_3}{d\zeta_p d\zeta_q d\zeta_l} \,.
\end{eqnarray}
Boosting this expression by an amount $\zeta_B=\tanh^{-1} V_r$, where $V_r$ is the transverse velocity, we find 
\begin{eqnarray}
{\tilde C}_3= \frac{\cosh\zeta_p\cosh\zeta_q\cosh\zeta_l}{\cosh(\zeta_p-\zeta_B)\cosh(\zeta_q-\zeta_B)\cosh(\zeta_l-\zeta_B)} C_3 \,, 
\label{eq:boost}
\end{eqnarray}
where we use a tilde to denote the boosted quantity.  The quantity $C_3$ is evaluated in the local rest frame and was given in eq.~(\ref{eq:C3r1})--expressed in terms of the single particle distribution, it is  given by eq.~(\ref{eq:C3r2}).  There have been simulations performed by a number of experimental groups \cite{Pruneau:2006gj,Pruneau:2007mq,Pruneau:2007ua,Ulery:2006ix}.  A first measurement of three particle azimuthal correlations has recently been reported by the STAR collaboration~\cite{:2008nd}. We expect the relevant quantity to be the three particle cumulant, $C_3$ divided by the cubic power of the single particle distribution. In addition, it is much more relevant to discuss the correlation between the relative angles of the particles.  

We now write the three particle correlation normalized by the cubic power of single particle distribution expressed in terms of the relative angles $\Delta\phi_{pq}\equiv \phi_p-\phi_q$ and $\Delta\phi_{pl}\equiv \phi_p-\phi_l$:
\begin{eqnarray}
\frac{{\tilde C}_3(\Delta\phi_{pq},\Delta\phi_{pl})}{C_1C_1C_1(\Delta\phi_{pq},\Delta\phi_{pl})}\equiv\frac{\kappa_3}{S_\perp^2 Q_S^4} \mathcal{A}(\Delta\phi_{pq},\Delta\phi_{pl},\zeta_B)
\end{eqnarray}
The explicit expression for $\mathcal{A}(\Delta\phi_{pq},\Delta\phi_{pl},\zeta_B)$ is given in eq.~(\ref{eq:A}) of Appendix C.  It is an analytical function of the relative angles and the transverse velocity.  It is normalized such that, for $\zeta_B=0$, we have $\mathcal{A}(\Delta\phi_{pq},\Delta\phi_{pl})\equiv 1$.

\begin{figure}
\centering
\hbox{
\includegraphics[scale=.7]{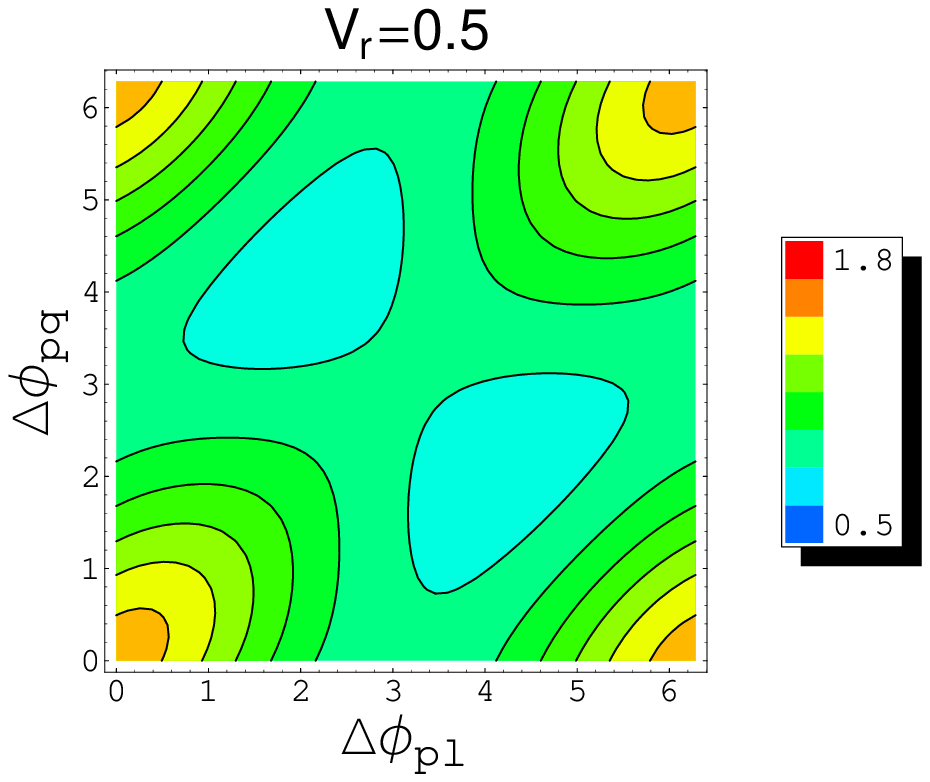}
\includegraphics[scale=.7]{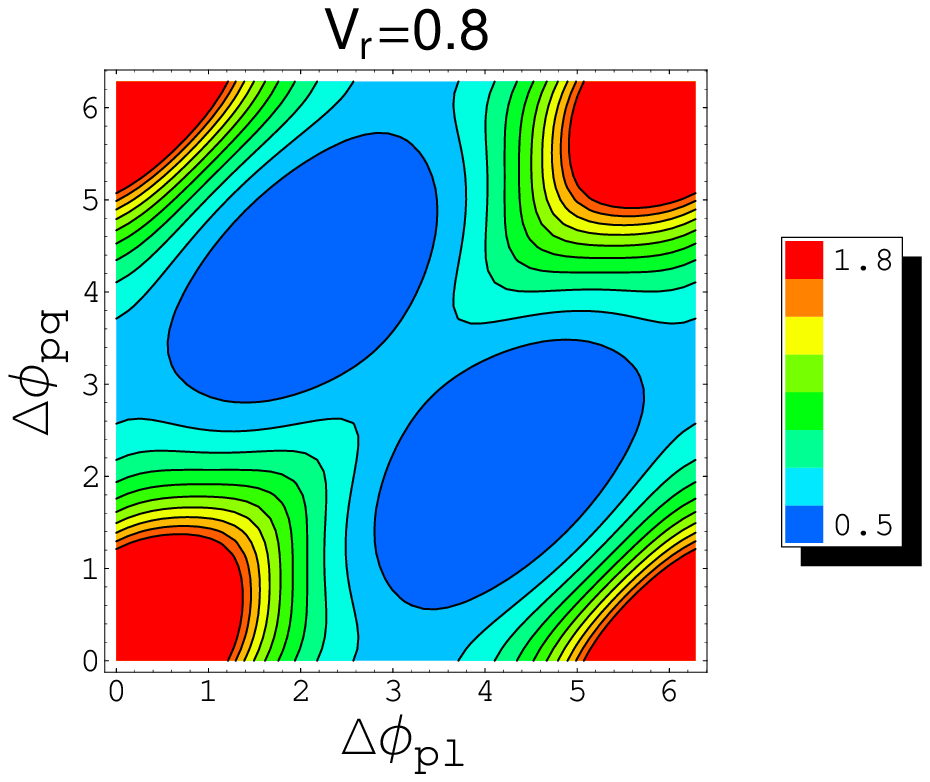}
}
\caption{(Color online) Contour plot of $\mathcal{A}(\Delta \phi_{pq},\Delta\phi_{pl})$ for $V_r=0.5$ (left) and $V_r=0.8$ (right). This 
plot only depicts the azimuthal structure of ridge like three particle near side correlations and does not take into account near side-- 
away side correlations arising from jet fragmentation.}
\label{fig:angcorr}
\end{figure}

The function $\mathcal{A}$ is plotted in figure~\ref{fig:angcorr} for radial flow values $V_r=0.5$ and $0.8$.  The radial flow collimates the angular distribution of particles in the direction of the flow. This is clearly seen in our results; we find that maximum amplitude always occurs at 
$\Delta\phi_{pq}=\Delta\phi_{pl}=0$--all three particles emitted colinearly. This feature of our result is consistent with first experimental observations of 
three particle azimuthal correlations~\cite{:2008nd}. As a corollary to this result, regardless of the magnitude of the boost, we find a minimum for $\Delta\phi_{pq}=2\pi/3$ and $\Delta\phi_{pl}=4\pi/3$ corresponding to the three particles being emitted with the furthest possible angular separation.  As the boost velocity is increased, the ratio between the maximum and minimum correlation increases. To observe these, the strength of the away side correlation coming from other  ``jet-like" mechanisms have to be subtracted. 

\section{Discussion}
\label{sec:discussion}

In the previous sections, we described the consequences of a novel mechanism for multiparticle production for long range three particle 
correlations in heavy ion collisions. The underlying picture is quite simple. Boost invariant Glasma flux tubes of transverse size $1/Q_S$ are formed in 
heavy ion collisions. They give the leading contribution to particle production. (For large rapidity separations, there can be significant violations of 
boost invariance--these too can be computed in the Glasma formalism~\cite{Gelis:2008sz}.) Particles are produced with equal probability along the length of the flux tube and decay isotropically. The flux tubes flow radially outwards due to the strong pressure gradients generated in the collision. While this 
does not significantly affect long range rapidity correlations, the azimuthal distributions are significantly altered because the emitted particles are collimated 
in the direction of the radial flow. In Refs.~\cite{Dumitru:2008wn,Gavin:2008ev}, this picture was shown to provide a quantitative description of STAR data 
on the near side ridge. Three particle correlations provide an additional measure to test this picture.  They predict that the strength of the correlation is 
proportional to $1/N_{\rm part}^2$. 

\begin{figure}
\centering
\includegraphics[scale=.75]{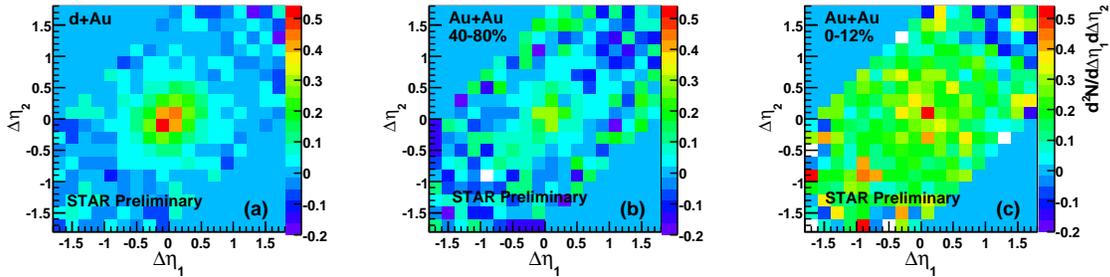}
\caption{(Color online) Preliminary results from STAR \protect\cite{Netrakanti:2008jw} of the background subtracted three particle pseudo-rapidity correlation in (left to right) d+Au, 40-80\% Au+Au and 0-12\% Au+Au collisions. The $p_\perp$ cuts for the trigger and associated particles are $3< p_\perp < 10$ GeV and $1 < p_\perp < 3$ GeV respectively. }
\label{fig:data1}
\end{figure}

In the above classical computation, the resulting three particle correlation has equal strength at all rapidity. The above calculation therefore predicts a structure-less correlation function of finite amplitude in $\Delta \eta_{pq}$ vs. $\Delta \eta_{pl}$.  Preliminary observations of such an effect have already been made \cite{Netrakanti:2008jw} and shown in fig.~\ref{fig:data1}.  A near-side ridge phenomenon is seen in the three particle correlation as one goes to more central collisions.  In the right most plot of fig.~\ref{fig:data1} a clear jet like structure is seen sitting on top of a structureless plateau that extends as far as 1.5 units in rapidity.  

\begin{figure}
\centering
\includegraphics[scale=1]{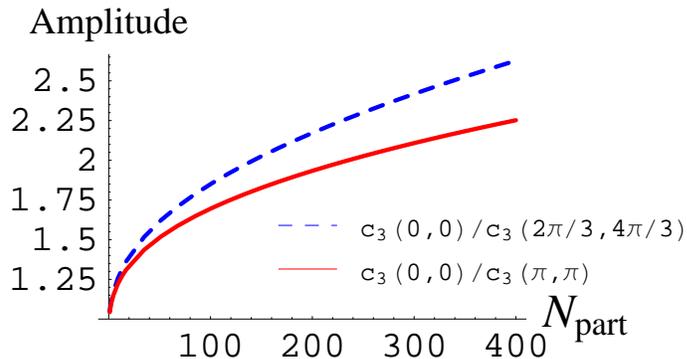}
\caption{Ratio of the maximum correlation $C_3(\Delta\phi_{pq}=0,\Delta\phi_{pl}=0)$ versus the minimum located at $C_3(\Delta\phi_{pq}=2\pi/3,\Delta\phi_{pl}=4\pi/3)$ versus centrality shown as the dashed blue curve.  The solid red curve shows the ratio at $\Delta\phi_{pq}=\Delta\phi_{pl}=\pi$ for comparison.  }
\label{fig:npart}
\end{figure}

It is the transverse boost that yields non-trivial angular correlations.  The PHENIX collaboration \cite{Kiyomichi:2005va} and the STAR collaboration~\cite{:2008ez} have extracted an average transverse velocity as a function of $N_{\rm part}$ from blast wave fits to data.  We use the PHENIX result as input into our boosted flux tube model.  We found in the previous section that a maximum correlation occurs for $\Delta \phi_{pq}=\Delta\phi_{pl}=0$ while a minimum occurs at $\Delta\phi_{pq}=2\pi/3$ and $\Delta\phi_{pl}=4\pi/3$.  It is therefore instructive to plot the ratio $C_3(0,0)/C_3(2\pi/3,4\pi/3)$ versus $N_{\rm part}$ as shown in fig.~\ref{fig:npart}.  We have chosen this ratio because it eliminates the explicit dependence on $\alpha_s$, $\kappa_3$ and $\mbox{N}_{\rm part}^2$.  Therefore, the only remaining dependence is on the radial flow $\zeta_B$ as a function of centrality.  We expect to see this behavior of the amplitude irrespective of the rapidity gap between particles. It can therefore be used as a test of our model.  Alternatively, by fitting the ratio of the three particle correlation, one can extract an independent determination of the transverse flow of the system.   
We should also note that the recent experimental data on three particle azimuthal correlations~\cite{:2008nd} does not show ``cross-talk" term that would 
show up as horizontal and vertical strips in the contour plot of fig.~(\ref{fig:angcorr}). This result is also consistent with the Glasma flux tube picture because, in the latter, azimuthal correlations are produced by angular collimation of the ridge particles by radial flow--this does not permit ``cross-talk" terms. 
Such terms are instead expected in scenarios where the angular correlation is sensitive to a ridge formed by jet-medium interactions.

\section{Conclusion}

We computed three particle correlations emanating from Glasma flux tubes formed in the collision of two sheets of Colored Glass Condensate.  A simple geometric interpretation of the strength of the correlations is found; the correlation is a number of order unity times the square of the ratio of the flux tube size over the system size, or equivalently, it is inversely proportional to the square of the number of participants.  
The correlation is {\em flat} in $\Delta\eta_{pq},\Delta\eta_{pl}$.  Non-trivial azimuthal correlations develop due to the 
radial transverse flow of particles produced by the flux tubes.  Finally, we predicted the relative strength of the correlation as a function of centrality using a blast wave model. This ratio removes the uncertainty from the overall non-perturbative coefficient in our result.  It therefore lends itself as a method to independently extract the average radial flow of the system. Further improvements to our model computation include non-perturbative solutions of the classical Yang--Mills equations to compute the three particle correlations, 
a better hydrodynamic treatment of  radial flow and hadronization effects, and finally, proper treatment of the quantum effects that modify the leading order 
boost invariant treatment of multiparticle production in heavy ion collisions. In particular, combining the Glasma flux tube picture with hydrodynamical 
simulations is very important for understanding the transverse momentum dependence of two particle near side correlations. An interesting first attempt 
combining the Glasma flux tube picture with hydrodynamical evolution has appeared recently~\cite{Takahashi:2009na}.

\section*{Acknowledgements}
We would like to thank Adrian Dumitru, Sean Gavin, Francois Gelis, Tuomas Lappi, Ron Longacre, Larry McLerran, Lanny Ray, Claude Pruneau, Paul Sorensen, Peter Steinberg, Jun Takahashi and Fuqian Wang for very useful discussions. This manuscript was authored under DOE Contract No.~\#DE-AC02-98CH10886. DFF thanks the hospitality of the Nuclear Theory Group of BNL during his visit in 2008 when part of this work carried out.  He also acknowledges the financial support from the Spanish research projects FPA2004-02602, FIS2008-01323, UCM-CAM 910309, FPA2007-29115-E and from the FPI programme (BES-2005-6726).

\appendix
\section{Computation of diffractive contribution}

In this appendix,  we shall explicitly compute the diagram in fig.~\ref{fig:d1}.  After performing the contraction among the sources of this graph we are left with
\begin{eqnarray}
&&\mathcal{F}^{(1)}=(2\pi)^{12} \mu_A^{12}\delta_{ci}\delta_{ek}\delta_{gm}\delta_{fd}\delta_{bj}\delta_{hl}\delta(\kpn{5}-\kpn{6})\delta(\kpn{3}-\kpn{4})\delta(\kpn{1}-\kpn{2})\nonumber\\
&\times&\delta(\lp-\kpn{5}+\qp-\kpn{3})\delta(\pp-\kpn{2}+\lp-\kpn{6})\delta(\pp-\kpn{1}-\qp+\kpn{4})
\end{eqnarray}
Upon substituting the above expression into eqn.~(\ref{eq:c}) we find\footnote{In evaluating the expression below we have used the convienent property that $L^2(\pp,\pp-\kpn{1})=L^2(\pp,\kpn{1})$}
\begin{eqnarray}
C_3^{(1)}=-\frac{1}{8(2\pi)^{11}}\left(g^3\right)^6N_c^3(N_c^2-1)\mu_A^{12}S_\perp\int\frac{d^2\kpn{1}}{\kpn{1}^{12}}\frac{L^2(\pp,\kpn{1}) L^2(\qp,\kpn{1}) L^2(\lp,\kpn{1})}{(\kpn{1}-\pp)^4(\kpn{1}-\qp)^4(\kpn{1}-\lp)^4}\nonumber\\
\label{eq:c22}
\end{eqnarray}

We should note that in order to get the above expression into a form symmetric with respect to $\pp,\qp$ and $\lp$ we had to do a trivial shift in integration variables ($\kpn{1}\to \kpn{1}+\pp$).  We now make use of the scalar product of two Lipatov vectors 
\begin{eqnarray}
L_\alpha({\bf p},{\bf k}_\perp)L^\alpha({\bf p},{\bf l}_\perp)=-\frac{4}{{\bf p}_\perp^2}\left[\delta^{ij}\delta^{lm}+\epsilon^{ij}\epsilon^{lm}\right]{\bf k}_\perp^i\left({\bf p}_\perp-{\bf k}_\perp\right)^j {\bf l}_\perp^l\left({\bf p}_\perp-{\bf l}_\perp\right)^m
\end{eqnarray}
and after noting that any angular dependence present in numerator will cancel we are left with
\begin{eqnarray}
C_3^{(1)}=\frac{64}{8(2\pi)^{11}}\left(g^3\right)^6N_c^3(N_c^2-1)\mu_A^{12}S_\perp\int\frac{d^2\kpn{1}}{\kpn{1}^{6}}\frac{1}{(\kpn{1}-\pp)^4(\kpn{1}-\qp)^4(\kpn{1}-\lp)^4}\nonumber\\
\end{eqnarray}
As we are interested in the limit when $\pp,\kp,\lp \gg Q_S$ we keep the term with the fewest powers of $\kpn{1}$ in the denominator.  In this limit the above expression becomes 
\begin{eqnarray}
C_3^{(1)}=\frac{64}{8(2\pi)^{11}}\left(g^3\right)^6N_c^3(N_c^2-1)\mu_A^{12}S_\perp\frac{2\pi}{\pp^4\qp^4\lp^4}\int\frac{d|\kpn{1}|}{|\kpn{1}|^{5}}
\end{eqnarray}
In order to evaluate the above integral we must introduce an infrared cut-off $k_{\mbox{min}}\sim Q_S$.  $Q_S$ acts as a natural cut-off as it signifies the onset of non-linear contributions to the infrared gluon distributions in the CGC. The final result is 
\begin{eqnarray}
C_3^{(1)}=\frac{16}{8(2\pi)^{11}}\left(g^3\right)^6N_c^3(N_c^2-1)\mu_A^{12}S_\perp\frac{2\pi}{\pp^4\qp^4\lp^4}\frac{1}{Q_S^4}
\end{eqnarray}
We should note that sub-leading contributions in individual graphs may have significant angular correlations. Our conjecture about the generality of the 
geometrical picture that emerges at high transverse momenta relies on these angular dependences cancelling among the (many) sub-leading graphs, as 
well as being smeared out by non-linear rescattering corrections. 

\section{Computation of the interference contribution}

We shall here explicitly compute the diagram of fig.~\ref{fig:d2}.  After performing the contraction among the sources of this graph we are left with
\begin{eqnarray}
&&\mathcal{F}^{(2)}=(2\pi)^{12} \mu_A^{12}\delta_{ci}\delta_{ek}\delta_{gm}\delta_{fh}\delta_{dl}\delta_{bj}\delta(\kpn{5}-\kpn{6})\delta(\kpn{3}-\kpn{4})\delta(\kpn{1}-\kpn{2})\nonumber\\
&\times&\delta(\lp-\kpn{5}-\pp+\kpn{2})\delta(\qp-\kpn{3}-\lp-\kpn{6})\delta(\pp-\kpn{1}-\qp+\kpn{4})\, .
\end{eqnarray}
Substituting the above expression into eqn.~(\ref{eq:c}), we find
\begin{eqnarray}
C_3^{(2)}=-\frac{1}{8(2\pi)^{11}}\left(g^3\right)^6N_c^3(N_c^2-1)\mu_A^{12}S_\perp\int\frac{d^2\kpn{1}}{\kpn{1}^{12}}\frac{L^2(\pp,\kpn{1}) L^2(\qp,\kpn{1}) L^2(\lp,\kpn{1})}{(\kpn{1}-\pp)^4(\kpn{1}-\qp)^4(\kpn{1}-\lp)^4}\nonumber\\
\end{eqnarray}
which is the same as expression (\ref{eq:c22}). The subsequent steps are identical.

\section{Angular Integrations}
\label{sec:C}

In this appendix, we work out the angular integrations necessary for evaluating the three particle correlation after a boost in the transverse direction.

First, let us define the three particle distribution and the cubic power of the single particle distribution in terms of relative angles.
\begin{eqnarray}
{\tilde C}_3(\Delta\phi_{pq},\Delta\phi_{pl})\equiv\int {\tilde C}_3(\phi_p,\phi_q,\phi_l) \delta(\Delta\phi_{pq} - \phi_p+\phi_q)\delta(\Delta\phi_{pl}-\phi_p+\phi_l) d\phi_p d\phi_q d\phi_l \nonumber\\
C_1C_1C_1(\Delta\phi_{pq},\Delta\phi_{pl})\equiv \int C_1(\phi_p) C_1(\phi_q) C_1(\phi_l) \delta(\Delta\phi_{pq} - \phi_p+\phi_q)\delta(\Delta\phi_{pl}-\phi_p+\phi_l) d\phi_p d\phi_q d\phi_l\nonumber\\
\end{eqnarray}

Now, let us evaluate ${\tilde C}_3(\Delta\phi_{pq},\Delta\phi_{pl})$.  The azimuthal dependence is dictated by the Jacobian in eq.~(\ref{eq:boost}). Upon substitution into the previous expression for ${\tilde C}_3(\Delta\phi_{pq},\Delta\phi_{pl})$, we are left with the following integrals  
\begin{eqnarray}
\tilde{C}_3(\Delta\phi_{pq},\Delta\phi_{pl})=C_3\int\int\int&& \frac{\cosh\zeta_p\cosh\zeta_q\cosh\zeta_l}{\cosh(\zeta_p-\zeta_B)\cosh(\zeta_q-\zeta_B)\cosh(\zeta_l-\zeta_B)} \nonumber\\&&\times\delta(\Delta\phi_{pq} - \phi_p+\phi_q)\delta(\Delta\phi_{pl}-\phi_p+\phi_l) d\phi_p d\phi_q d\phi_l \nonumber\\
\end{eqnarray}
where $C_3$ is the local rest frame quantity computed in eqn.~(\ref{eq:C3r1}) or \ref{eq:C3r2}.  The above integrals can be done analytically and we can express the result as
\begin{eqnarray}
\tilde{C}_3(\Delta\phi_{pq},\Delta\phi_{pl})=2\pi C_3\mathcal{A}(\Delta\phi_{pq},\Delta\phi_{pl},\zeta_B)
\end{eqnarray}
where
\begin{eqnarray}
& &\mathcal{A}(\Delta\phi_{pq},\Delta\phi_{pl},\zeta_B)\nonumber\\&=&\left[25+36 \cosh(2\zeta_B)+3\cosh(4\zeta_B)-8(\cos\Delta\phi_{pq}+\cos(\Delta\phi_{pq}-\Delta\phi_{pl})+\cos\Delta\phi_{pl}) \sinh^4\zeta_B\right]\nonumber\\
&\times&[3+\cosh(2\zeta_B)-2\cos\Delta\phi_{pq}\sinh^2\zeta_B]^{-1}\nonumber\\
&\times&[3+\cosh(2\zeta_B)-2\cos\Delta\phi_{pl}\sinh^2\zeta_B]^{-1}\nonumber\\
&\times&[3+\cosh(2\zeta_B)-2\cos(\Delta\phi_{pq}-\Delta\phi_{pl})\sinh^2\zeta_B]^{-1}
\label{eq:A}
\end{eqnarray}
The angular integrals in the expression for $C_1C_1C_1(\Delta\phi_{pq},\Delta\phi_{pl})$ result in an overall normalization only,
\begin{eqnarray}
C_1C_1C_1(\Delta\phi_{pq},\Delta\phi_{pl})=2\pi\left<\frac{dN}{dy_pd^2{\bf p}_\perp}\right>\left<\frac{dN}{dy_qd^2{\bf q}_\perp}\right>\left<\frac{dN}{dy_ld^2{\bf l}_\perp}\right>
\end{eqnarray}
The final result is
\begin{eqnarray}
\frac{{\tilde C}_3(\Delta\phi_{pq},\Delta\phi_{pl})}{C_1C_1C_1(\Delta\phi_{pq},\Delta\phi_{pl})} = \frac{\kappa_3}{S_\perp^2 Q_S^4} \mathcal{A}(\Delta\phi_{pq},\Delta\phi_{pl},\zeta_B)
\end{eqnarray}


\begin{thebibliography}{MM}
\bibitem{STAR1}
J.~Putschke,J. Phys. {\bf G34}, S679 (2007);
M.~Daugherity, arXiv:0806.2121 [nucl-ex].
\bibitem{PHENIX}
A.~Adare {\it et al.}  [PHENIX Collaboration],
arXiv:0801.4545 [nucl-ex].
\bibitem{Alver:2008gk}
B.~Alver {\it et al.}  [PHOBOS Collaboration],  
  arXiv:0812.1172 [nucl-ex].
\bibitem{STAR2}
J.~Adams et al. [STAR Collaboration]
Phys.\ Rev.\ Lett. {\bf 95}:152301, (2005);  Fuqiang Wang [STAR Collaboration],
talk at Quark Matter 2004, J.\ Phys.\ G {\bf 30}:S1299-S1304, (2004).
\bibitem{Molnar:2007wy}
  L.~Molnar,
  J.\ Phys.\ G {\bf 34}, S593 (2007)
  [arXiv:nucl-ex/0701061].
\bibitem{Wenger:2008ts}
  B.~Alver {\it et al.}  [PHOBOS Collaboration],
  J.\ Phys.\ G {\bf 35}, 104080 (2008)
  [arXiv:0804.3038 [nucl-ex]].
\bibitem{STAR3}
J.~Adams {\it et al.}  [STAR Collaboration],
  Phys.\ Rev.\  C {\bf 73}, 064907 (2006).
\bibitem{Models}
N.~Armesto, C.~A.~Salgado, U.~A.~Wiedemann,
  Phys.\ Rev.\ Lett.\  {\bf 93}, 242301 (2004); P.~Romatschke,
  Phys.\ Rev.\  C {\bf 75}, 014901 (2007);  A.~Majumder, B.~Muller, S.~A.~Bass,
  Phys.\ Rev.\ Lett.\  {\bf 99}, 042301 (2007); C.~B.~Chiu, R.~C.~Hwa,
  Phys.\ Rev.\  C {\bf 72}, 034903 (2005); C.~Y.~Wong,
  arXiv:0712.3282 [hep-ph]; R.~C.~Hwa, C.~B.~Yang,
  arXiv:0801.2183 [nucl-th]; T.~A.~Trainor,
  arXiv:0708.0792 [hep-ph]; A.~Dumitru, Y.~Nara, B.~Schenke, M.~Strickland,
  arXiv:0710.1223 [hep-ph]; S.~J.~Lindenbaum,  R.~S.~Longacre,
  Eur.\ Phys.\ J.\  C {\bf 49}, 767 (2007).
\bibitem{Dumitru:2008wn}
  A.~Dumitru, F.~Gelis, L.~McLerran and R.~Venugopalan,
  Nucl.\ Phys.\  A {\bf 810}, 91 (2008)
  [arXiv:0804.3858 [hep-ph]].
\bibitem{Gavin:2008ev}
  S.~Gavin, L.~McLerran and G.~Moschelli,
  arXiv:0806.4718 [nucl-th].
\bibitem{Netrakanti:2008jw}
  P.~K.~Netrakanti  [STAR Collaboration],
  J.\ Phys.\ G {\bf 35}, 104010 (2008)
  [arXiv:0804.4417 [nucl-ex]]. 
\bibitem{CGC}
E. Iancu, R. Venugopalan, hep-ph/0303204.
\bibitem{MV}
L.McLerran and R. Venugopalan, Phys. Rev. {\bf D 49}, 2233 (1994).  {\it ibid}. {\bf D 49}, 3352 (1994); {\bf D 50}, 2225 (1994).
\bibitem{Gelis:2008rw}
  F.~Gelis, T.~Lappi and R.~Venugopalan,
  Phys.\ Rev.\  D {\bf 78}, 054019 (2008)
  [arXiv:0804.2630 [hep-ph]].
\bibitem{Gelis:2008ad}
  F.~Gelis, T.~Lappi and R.~Venugopalan,
  Phys.\ Rev.\  D {\bf 78}, 054020 (2008)
  [arXiv:0807.1306 [hep-ph]].
\bibitem{JIMWLK}
J. Jalilian-Marian, A. Kovner, L.D. McLerran, H. Weigert, Phys. Rev. {\bf D 55}, 5414 (1997);
{J. Jalilian-Marian, A. Kovner, A. Leonidov, H. Weigert}, Nucl. Phys. {\bf B 504}, 415 (1997);
{J. Jalilian-Marian, A. Kovner, A. Leonidov, H. Weigert}, Phys. Rev. {\bf D 59}, 034007 (1999);
{E. Iancu, A. Leonidov, L.D. McLerran}, Nucl. Phys. {\bf A 692}, 583
  (2001); {E. Ferreiro, E. Iancu, A. Leonidov, L.D. McLerran}, Nucl. Phys. {\bf A 703}, 489 (2002).
\bibitem{Gelis:2008sz}
  F.~Gelis, T.~Lappi and R.~Venugopalan,
  arXiv:0810.4829 [hep-ph].
\bibitem{Blaizot:2008yb}
  J.~P.~Blaizot and Y.~Mehtar-Tani,
  Nucl.\ Phys.\  A {\bf 818}, 97 (2009)
  [arXiv:0806.1422 [hep-ph]].
\bibitem{Kovner:1995ts}
  A.~Kovner, L.~D.~McLerran and H.~Weigert,
  Phys.\ Rev.\  D {\bf 52}, 3809 (1995)
  [arXiv:hep-ph/9505320].
\bibitem{Kovchegov:1997ke}
  Y.~V.~Kovchegov and D.~H.~Rischke,
  Phys.\ Rev.\  C {\bf 56}, 1084 (1997)
  [arXiv:hep-ph/9704201].
\bibitem{KNV}
A. Krasnitz, R. Venugopalan, Nucl. Phys. {\bf B 557}, 237 (1999);
Phys. Rev. Lett. {\bf 84}, 4309 (2000); {\it ibid.}, {\bf 86}, 1717 (2001); A. Krasnitz, Y. Nara, R. Venugopalan, Phys. Rev. Lett. {\bf 87}, 192302  (2001); Nucl.\ Phys.\  A {\bf 717}, 268 (2003); {\it ibid.}, {\bf A 727}, 427 (2003);
T.~Lappi,  Phys.\ Rev.\  C {\bf 67}, 054903 (2003).
\bibitem{LappiMcLerran}
T.~Lappi and L.~McLerran, Nucl.\ Phys.\  A {\bf 772}, 200 (2006).
\bibitem{GelisVLectures}
F.~Gelis, R.~Venugopalan, Acta Phys.\ Polon.\  B {\bf 37}, 3253 (2006); F.~Gelis, T.~Lappi, R.~Venugopalan, Int.\ J.\ Mod.\ Phys.\  E {\bf 16}, 2595 (2007).
\bibitem{KharzeevKV}
D.~Kharzeev, A.~Krasnitz and R.~Venugopalan, Phys.\ Lett.\ B {\bf 545}, 298 (2002).
\bibitem{KharzeevMcLW}
D.~E.~Kharzeev, L.~D.~McLerran and H.~J.~Warringa, Nucl.\ Phys.\  A {\bf 803}, 227 (2008).
\bibitem{Voloshin}
S.~A.~Voloshin, Phys.\ Lett.\  B {\bf 632}, 490 (2006).
\bibitem{Shuryak}
E.~V.~Shuryak,
  Phys.\ Rev.\  C {\bf 76}, 047901 (2007).
\bibitem{Gelis:2009wh}
  F.~Gelis, T.~Lappi and L.~McLerran,
  arXiv:0905.3234 [hep-ph].
\bibitem{LappiSrednyakRV}
T. Lappi, S. Srednyak and R. Venugopalan, in preparation.
\bibitem{Lappi:2007ku}
  T.~Lappi,
  Eur.\ Phys.\ J.\  C {\bf 55}, 285 (2008).
\bibitem{Gunion:1981qs}
  J.~F.~Gunion and G.~Bertsch,
  Phys.\ Rev.\  D {\bf 25}, 746 (1982).
\bibitem{Gyulassy:1997vt}
  M.~Gyulassy and L.~D.~McLerran,
  Phys.\ Rev.\  C {\bf 56}, 2219 (1997)
  [arXiv:nucl-th/9704034].
\bibitem{Pruneau:2006gj}
  C.~A.~Pruneau,
  Phys.\ Rev.\  C {\bf 74}, 064910 (2006)
  [arXiv:nucl-ex/0608002].
\bibitem{Pruneau:2007mq}
  C.~Pruneau,
  Int.\ J.\ Mod.\ Phys.\  E {\bf 16}, 1964 (2007)
  [arXiv:nucl-ex/0703009].
\bibitem{Pruneau:2007ua}
  C.~A.~Pruneau, S.~Gavin and S.~A.~Voloshin,
  Nucl.\ Phys.\  A {\bf 802}, 107 (2008)
  [arXiv:0711.1991 [nucl-ex]].
\bibitem{Ulery:2006ix}
  J.~G.~Ulery and F.~Wang,
  arXiv:nucl-ex/0609017.
 \bibitem{:2008nd}
  B.~I.~Abelev {\it et al.}  [STAR Collaboration],
  Phys.\ Rev.\ Lett.\  {\bf 102}, 052302 (2009)
  [arXiv:0805.0622 [nucl-ex]]. 
\bibitem{Kiyomichi:2005va}
  A.~Kiyomichi  [PHENIX Collaboration],
{\it Prepared for Lake Louise Winter Institute: Fundamental Interactions, Lake Louise, Alberta, Canada, 20-26 Feb 2005}.
\bibitem{:2008ez}
  B.~I.~Abelev {\it et al.}  [STAR Collaboration],
  Phys.\ Rev.\  C {\bf 79}, 034909 (2009)
  [arXiv:0808.2041 [nucl-ex]].
  \bibitem{Takahashi:2009na}
  J.~Takahashi, B.~M.~Tavares, W.~L.~Qian, F.~Grassi, Y.~Hama, T.~Kodama and N.~Xu,
  arXiv:0902.4870 [nucl-th].
\end{thebibliography}
\end{document}